# Finite element modeling of slow water filtering

[1]Zhanna Alsar, [2]Bayan Kurbanova, [1]Nurzhan Serik, [1]Kurbangali Tnyshtykbayev, [3]Aisarat Gadzhimuradova, [1]Ainur Baikadamova, [4]Issakul Tumanov, [5]Z.A. Mansurov, [1, 6]Z. Insepov*

1 Satbayev Kazak National Technical University, Satpaev 22, Almaty, RK
2 Nazarbayev University, Kabanbay Batyr 53, Astana, RK
3 Kazakh State Agrotechnical University, Zhengis Ave 62, Astana, RK
4 Al-Farabi Kazakh National University, Al-Farabi St. 71 Almaty RK
5 Institute of Combustion Problems, Masanchi St 53, Almaty, RK
6 Purdue University, School of Nuclear Engineering, West Lafayette, Indiana, USA
Corresponding author: zinsepov@purdue.edu

**ABSTRACT**

Slow sand filtration is most appropriate where there is funding to subsidize the initial cost of the filter, available training for use and maintenance, locally available sand, and a transportation network capable of moving the filter. Since the experimental study of slow filtration is taking a long time, theoretical analysis, modelling and simulation studies become important. Theoretical models of slow filtration are not yet sufficiently developed. Several reasons are hindering modelling and simulation: due to the complexity of the interaction of water and pollution with filters and the lack of detailed values for the kinetic coefficients in elementary processes. Multiphysics modelling using Comsol, where a complex set of Navier-Stokes equations is numerically solved together, with the molecular awareness for the kinetic coefficients, was used to studying various geometries and construction types of slow filtration.

**INTRODUCTION**

The goal of the paper: provide scientific substantiation and development of effective technological processes, technologies for purification of natural waters and improvement of the quality of drinking water in the regions of Kazakhstan.

The main efforts are to address theory, modelling and simulation of the slow filtration technology at low-capacity facilities for natural water purification.

Slow filtration technology was originally applied in the first half of the 19th century ([1], Chelsea, 1829). At a time when surface waters weren't heavily polluted and standards weren't as stringent, slow filtration seemed like the ideal "ecological" treatment, mimicking the natural way spring water was produced, because the water was filtered through the filter media without any added chemicals.

The company has built two reservoirs in Green Park, near Buckingham Palace. A third reservoir was built in Hyde Park. The water came from the River Thames. However, the river proved to be an unstable source of good drinking water. Increasing industrialization in 19th century polluted the river as sewers and factories dumped a constant stream of waste into its waters.

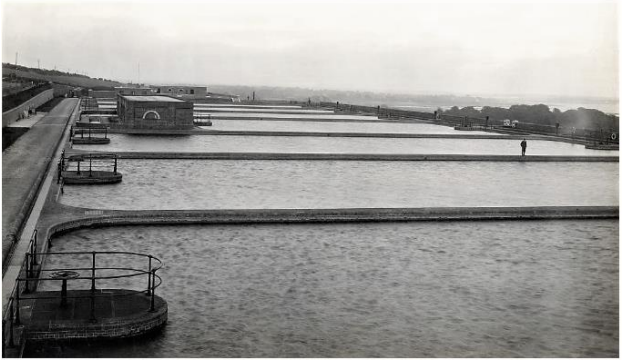

Figure 1 - Large-scale slow filtering. Portsmouth, 1927

The basic principle of the slow filtration process is simple. Polluted water flows through a layer of sand, where it is not only physically filtered, but also biologically treated. This removes deposits and pathogens. This process is based on the ability of microorganisms to remove disease-causing microbes.

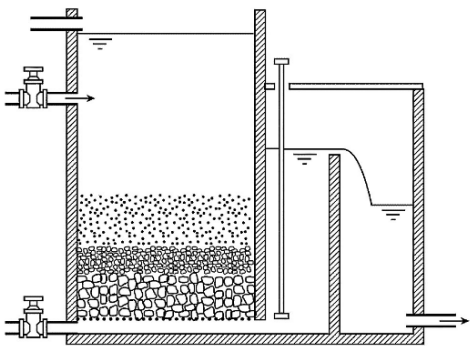

Figure 2 - The principle of a slow sand filter. Source: WHO.

In the treatment process, the raw water underwent only physical pre-treatment (micro-deformation or rapid pre-filtration through gravel and/or sand in "pre-filters", at a rate of approximately 5 m/d. Modern fast filters operate 30-50 times faster.

Slow filtration sometimes uses only bacteria. However, it often happens that after a certain period of maturation, a complex biocenosis arises on the surface of the sand, consisting of algae, bacteria and zooplankton. The latter acts simply as a predator and has a limiting effect on the first two. In this environment (the biological membrane) a complex symbiosis occurs between algae and bacteria similar to that found in a natural lagoon.

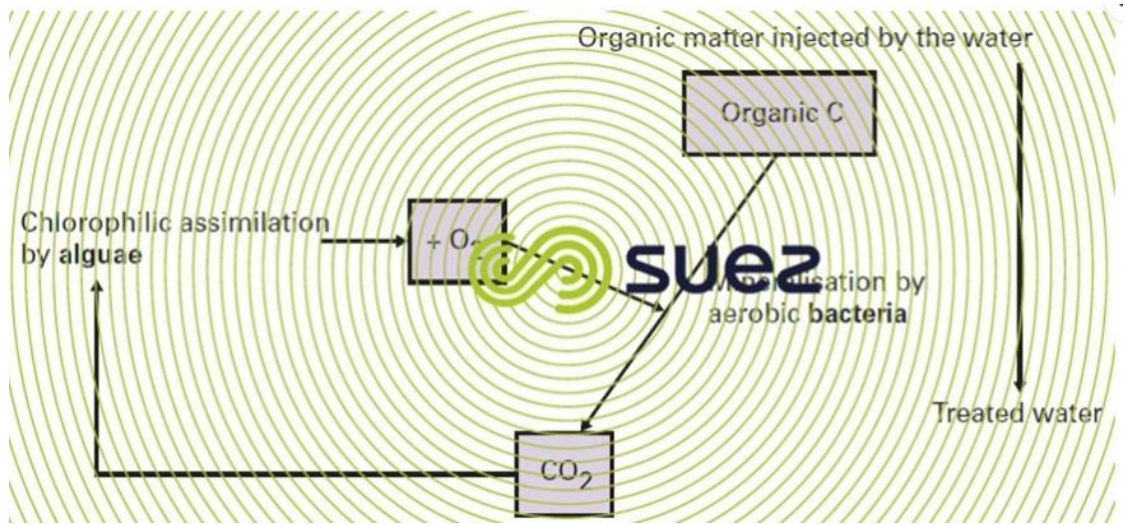

Figure 3 - Symbiosis of algae and bacteria in a slow filtration membrane.

The various modes of slow filtration biomass can be summarized as follows: mechanical retention and filtering effect in two successive media: the biological membrane and the actual sand; bioflocculation using exopolymers secreted by algae and bacteria; bacterial oxidation due to dissolved $O_2$ present in raw water and $O_2$ produced by algae; nitrogen-ammonia nitrification, mineralization of biodegradable organic matter; binding (heavy metals), bioaccumulation (detergents, pesticides), metabolism (phenols, pesticides).

Elimination of pathogenic bacteria through the "fight for life" system, bioflocculation, predation, antibiosis (some green algae, such as Chlorella, are able to secrete bactericidal substances).

The pressure loss changes very slowly due to the low filtration rate, and on average the filters are washed once a month.

Rinsing is usually done manually (using pressurized water jets) and sometimes mechanized (descaling). After the wash is completed, the filtered water will not always be satisfactory and will have to be drained down the drain until the biological membrane is restored. This process takes several days.

Slow filtration clarifies water well at low suspended solids content in raw water. As the solids content increases, the pre-treatment systems, gratings and pre-filters become inadequate and the turbidity of the treated water may well exceed the limit set by the standards.

In addition, these filters are sensitive to high levels of planktonic algae in raw water (surface contamination). Some species have been shown to create particularly high levels of pollution (eg Asterionella diatom and Pediastrum green algae, in microorganisms for which fresh water is their natural habitat).

Disadvantages of slow sand filtration:
- Not fully effective against viruses
- Lack of residual protection - may lead to re-infection.
- Regular cleaning can damage the biolayer and reduce efficiency.
- Difficult to transport due to weight - high initial cost

Slow sand filtration is most appropriate where there is funding to subsidize the initial cost of the filter, available training for use and maintenance, locally available sand, and a transportation network capable of moving the filter.

Theoretical models of slow filtration are not sufficiently developed. First of all, due to the complexity of the interaction of water and pollution with filters and the lack of detailed values for the kinetic coefficients in elementary processes.

Multiphysics modelling using Comsol gives a powerful tool to researchers for studying various geometries and construction types for slow filtration.

**Main parameters and governing equations.**

1. *Darcy's law.*

The flow in porous materials is characterized by the Darcy's law, whereas the linear relationship between the velocity field *u* (m/s) and the gradient of the pressure *p* (Pa) can be written as follows [2], (Darcy 1856):

$$u = -\frac{k}{\mu}\nabla p \qquad (1)$$

with *k* (m²) the permeability of the porous medium and *μ* (Pa·s) the dynamic viscosity of the fluid, respectively.

The permeability can be obtained from the Kozeny-Carman relationship for the packed beds or granular soils [3], (Kozeny 1927):

$$k = \frac{d_p^2}{180}\frac{\epsilon_p^3}{(1-\epsilon_p)^2} \qquad (2)$$

Here, $d_p$ (m) is the average granular diameter. Darcy's linear flow is only valid for very low velocities or at low Reynolds numbers (Re < 10).

## 2. Transport of dilute species in porous media.

This transport model is a sub-model of the physical model including transport of chemical species. It is used for the study of solute transport in porous media for the Darcy's law. The solute transport is directly related to the nature of the porous media and model includes reaction rate expressions and solute sources. This interface used to compute the concentration and transport of species in a free and porous media [4], (Comsol Interface).

$$\rho_{1j}\frac{dc_i}{dt} + \rho_{2i} + \nabla \cdot \Gamma_i + u \cdot \nabla c_i = R_i + S_I \qquad (3)$$

$$N_i = \Gamma_i + uc_i = -D_{e,i}\nabla c_i + uc_I \qquad (4)$$

where *ρ* is the bulk density, *c* is the concentration, *t* is time, *u* is the fluid velocity, *R* is the retardation factor, *S* is the saturation factor, *D* is the hydrodynamic dispersion coefficient, *N* is the average pore water velocity, and *Γ* is the gamma function.

**Numerical solution.**

Using the COMSOL Multiphysics software, we solved contaminated water transfer through a porous sorbent and ceramic medias with the porosities of 70% where the linear Darcy flow is dominating the flow.

This model is based on the ceramic water filter model [5], (Comsol works), which contained activated carbon core as sorbent and we have modified it under our problem to show the efficiency of the sorbent for water filtration. Sorbent has a large surface area and is mostly used in granular form for water filters. The advantage of granular sorbent compared to powdered-sorbent is the smaller pressure drop due to its relatively large particle size.

At the same time this results in a smaller surface area available for reaction and adsorption. Average sorbent granule size was varied from 0.03 to 0.3 mm. For our simulations, we assume chlorine and mercury as the solutes and water as the solvent. The use of sorbent for water filtration is based on the principle of adsorption, where contaminants are adsorbed onto the surface of the adsorbents. Adsorption rates of Chlorine was modeled using Freundlich adsorption isotherm for sorbent [6], (Alsar 2020). Freundlich constant and exponent for Mercury adsorbed at the surface of sorbent were extracted from the Freundlich isotherm curve, obtained in the present work. The curves were plotted with the help of Mercury porosimeter measurements' results. Also, we consider displacing fluid is pure water. The concentration of displaced fluid is calculated from the initial concentrative properties of aqueous solution. Transport properties used in our simulations are given in Table 1.

**Table 1. Contaminant transport properties used in this simulation**

| Name | Value | Description, refs |
| --- | --- | --- |
| K_f_ Cl | 0.1819 | Freundlich constant, Cl [6] |
| N_f_ Cl | 0.94 | Freundlich exponent, Cl [6] |
| Conc_Cl | 0.5 mol/$m^3$ | Reference concentration, Cl [6] |
| K_f_Hg | 0.03 | Freundlich constant, Hg [this work] |
| N_f_Hg | 1.25 | Freundlich exponent, Hg [this work] |
| Conc_Hg | 0.5 mol/$m^3$ | Reference concentration, Hg [this work] |

The computational domain was a three-dimensional (3D) as shown in Figure 1. The thickness of the sorbent and ceramic layers are 0.4 and 0.5 meters respectively, whereas filter's overall length and radius are 15 and 6 meters. Dirty water containing different impurities enters the filter with an inlet pressure of 6 psi. The inlet boundary condition specifies species concentration and outlet boundary condition corresponds to free flow.

First, the water passes through the porous ceramic medium, in which particles with a larger diameter than the pore size are filtered out. Then the remained species are entered through the sorbent layer and adsorbed onto the surface. Eventually, pure water without any concentrations can be obtained at the output flow.

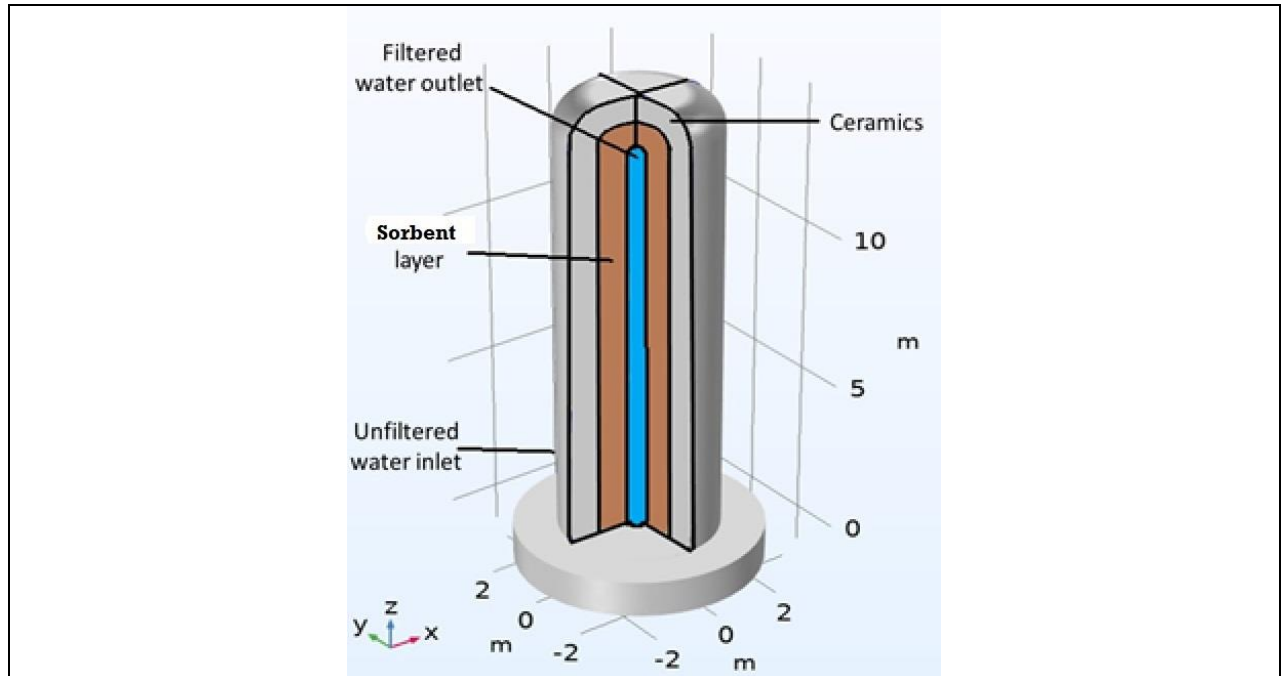

Figure 1. The computational domain, dimensions are in [m]. The sorbent material layer was used having detailed kinetic parameters for the simulation.

Figure 2 shows a hybrid structured grid used for the calculations of flow simulations in the porous sorbent and ceramic media where ultra-fine triangular mesh cells have been used near solid walls to provide better resolution and spread along with the computational domain.

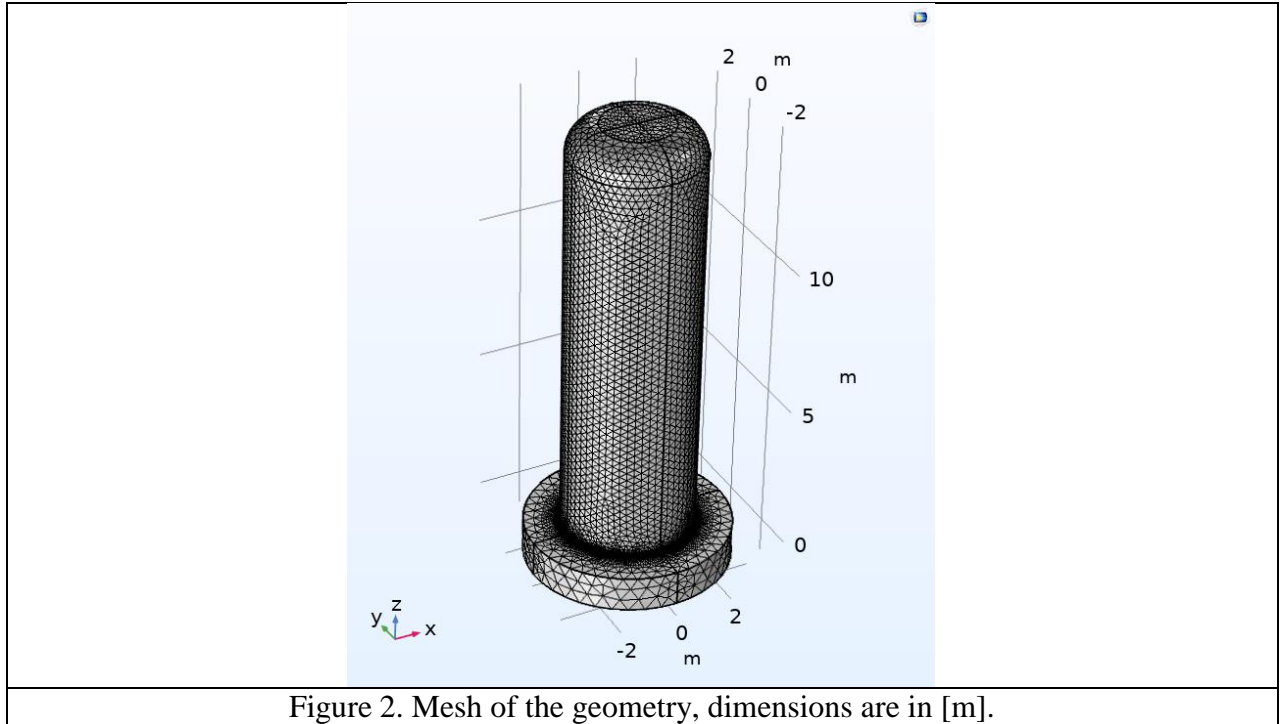
Figure 2. Mesh of the geometry, dimensions are in [m].

**Results and discussion**

The efficiency of the filters mainly depend on the ability of the adsorbent to adsorb the contaminants in the water sample and flow velocity of the liquid through porous structures. As sorbent has unique sorption abilities, it is essential to study the filtration process of different contaminants within various times. A time-dependent study was chosen to determine the change over time in the concentration of contaminants adsorbed by the sorbent and ceramic layers. The three-dimensional (3D) plots and line graphs as post-processing documents from Comsol Multiphysics simulation are shown in Figures below. Figure 3 demonstrates the variations of pressure and velocity field in the filter within 2 hours in different granular sized (0.3 and 0.03 mm) sorbent layers. The decrease of granular size leads to the overall increase and uniform distribution of the pressure and flow field inside the filter.

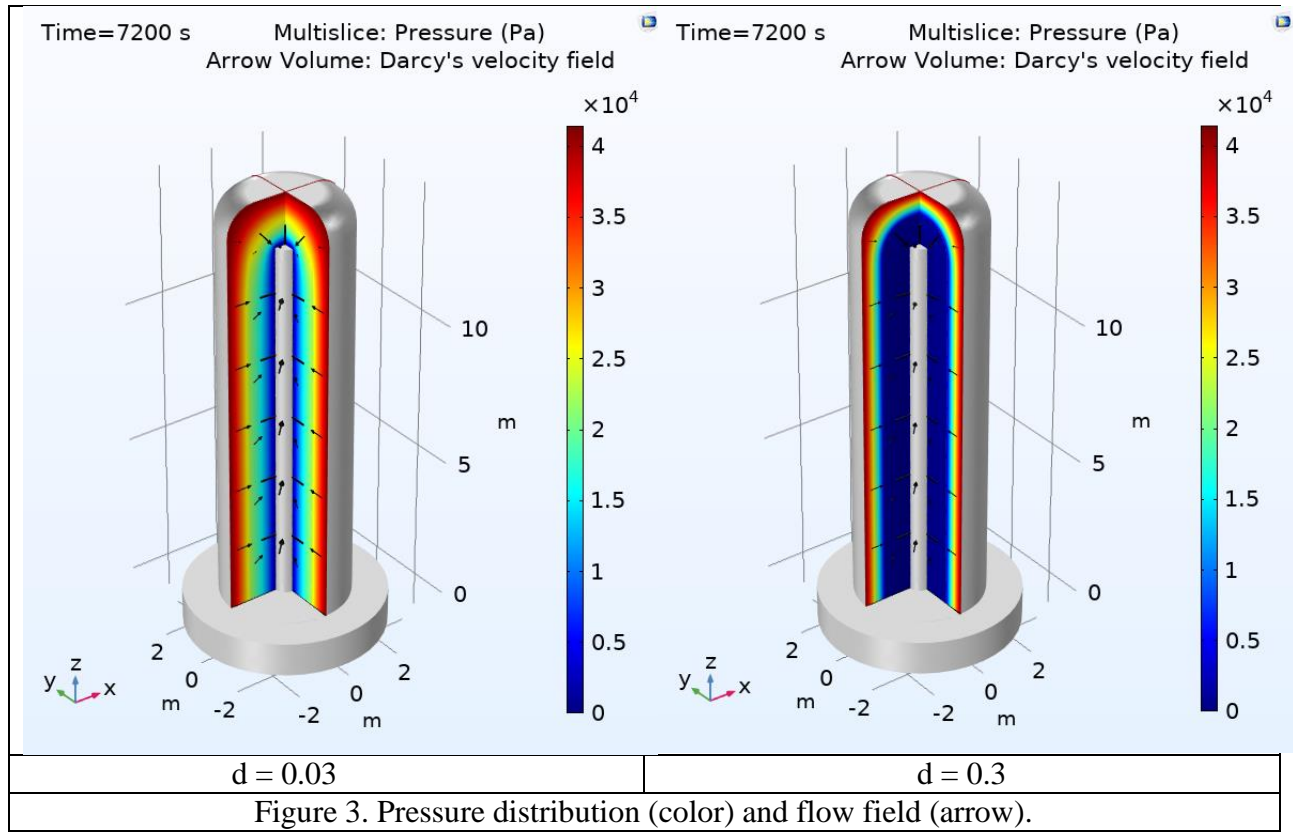

| d = 0.03 | d = 0.3 |

Figure 3. Pressure distribution (color) and flow field (arrow).

The variations of the adsorbed contaminant concentrations along the thickness of the ceramic and sorbent layers were monitored and examined. This was done by recording the output concentration and plotting the variations for the two hours. The surface concentration plots are showed a clear change in initial concentration as the diluted species flow through the ceramic and sorbent layers' thickness. Figure 4 demonstrates the Mercury concentration in the filter after 2 hours for 0.03 and 0.3 mm sized granular sorbent. There is a slight difference in mercury concentration along the filter between 0.03 and 0.3 mm granulas. Overall concentration is lower in small granular sized (0.03mm) filters than those with the large sized ones (0.3mm). However, after mercury has passed the ceramic part, it is removed from the water by the reaction with sorbent layer in both cases.

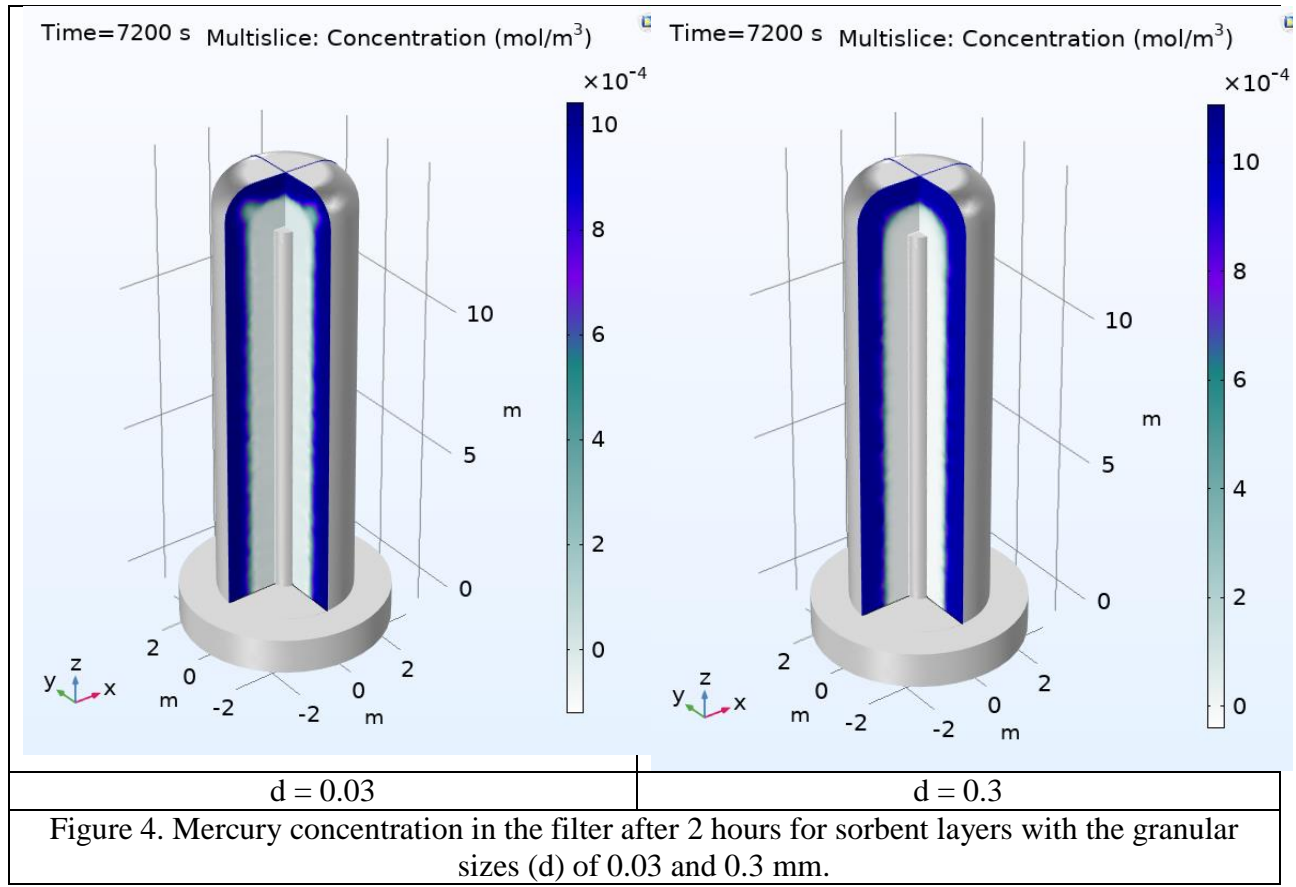

| d = 0.03 | d = 0.3 |

Figure 4. Mercury concentration in the filter after 2 hours for sorbent layers with the granular sizes (d) of 0.03 and 0.3 mm.

The concentrations of mercury in the filter for the various times were recorded and plotted against the corresponding times, and the results are as shown in Figure 5. It shows a significant approximation of the time intervals each layer will continue to give 100 percent efficiency, as the lifetime of the contaminant in the water is varied. It is observed that there might be incomplete removal of mercury in the output water filtered over a short period of time (5 minutes). If the filter is examined over a long period of time, the mercury clog the pores of the ceramic, which leads to the increase of mercury concentration in the ceramic layer. This might be attributed to the decrease in free adsorption sites for contaminant adsorption with time, due to the continuing adsorption of the contaminants onto the adsorbent surface. However, it can be seen that there is a complete deletion of the mercury at the output for a long filtration time.

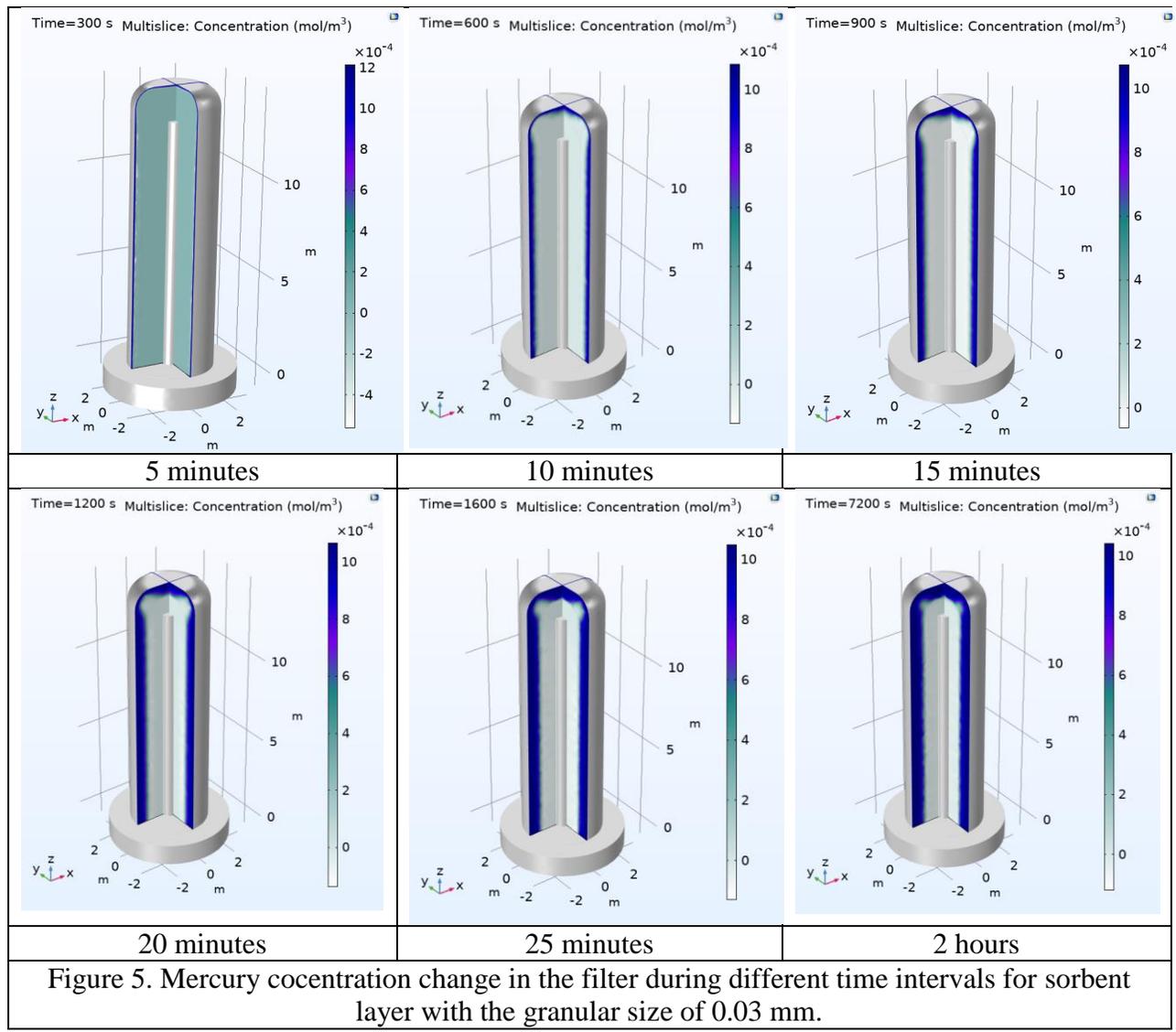

Figure 5. Mercury cocentration change in the filter during different time intervals for sorbent layer with the granular size of 0.03 mm.

It is interesting to note that the complete removal of the chlorine concentration doesn't depend on sorbent granular size as well as different filtration times. Comparing mercury, chlorine can be easily disposed of even at large sorbent granular sizes and short filtration times (Figures 6, 7).

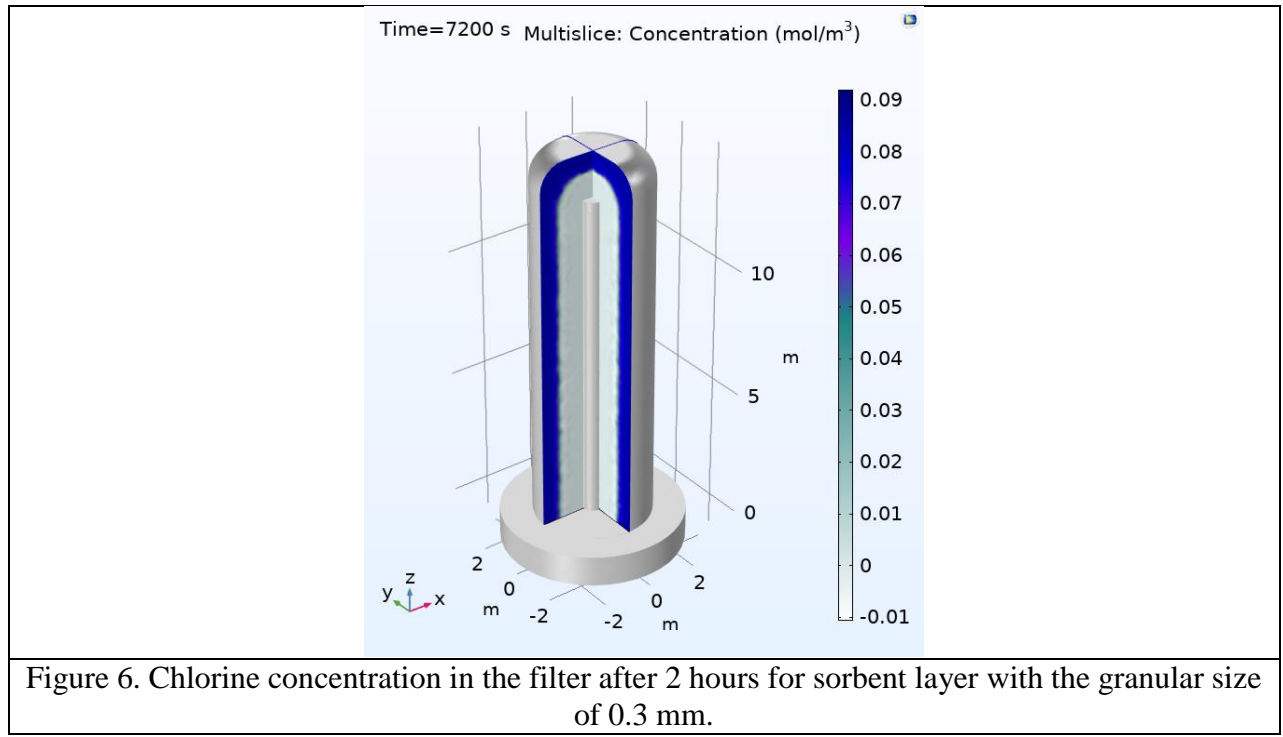

Figure 6. Chlorine concentration in the filter after 2 hours for sorbent layer with the granular size of 0.3 mm.

The line graphs of Figure 7 show a non-linear change in the output concentration of mercury adsorbed along the adsorbents thicknesses for a time range of 0 to 5 minutes, with intervals of 100 seconds. It is observed that, due to the different sizes of sorbent granule (0.03 and 0.3 mm), the height of the graph also varies accordingly.

The overall output concentration of mercury is lower in the filter with the small sized granular sorbent than the large sized ones. This may further explain the low efficiency of the large sized sorbent layers for complete removal of mercury in the filter. Moreover, there is a plateau in the mercury concentration at approximately 200 seconds. Further, the concentration is decreased continuously with the rising of filtration time. Having said that, it can be concluded that not only granular size, also filtration time is vital for the complete removal of the mercury and increase the whole efficiency of the filter.

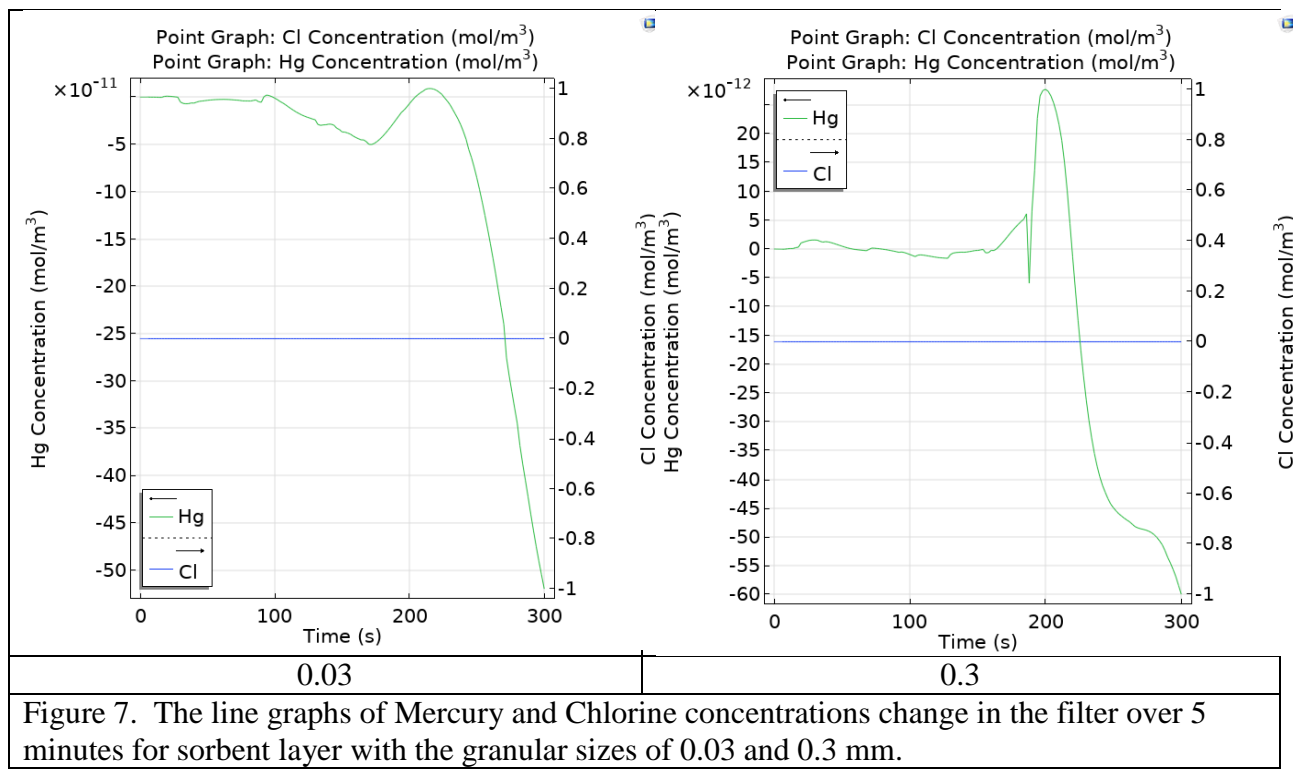

| 0.03 | 0.3 |

Figure 7. The line graphs of Mercury and Chlorine concentrations change in the filter over 5 minutes for sorbent layer with the granular sizes of 0.03 and 0.3 mm.

**CONCLUSIONS**

Theoretical models of slow filtration are not yet fully developed. Several reasons are hindering modelling and simulation. Due to the complexity of the interaction of water and contamination with sorbents and filters and the lack of detailed kinetic coefficients of the elemental processes. In this paper, the kinetic coefficients were obtained from experiment in this work and in Ref.[6]. Multiphysics modelling using Comsol, where a complex set of Navier-Stokes equations is numerically solved together, with the atomistic awareness of the kinetic coefficients, was used to studying various geometries and construction types of slow filtration. Future work would include partially or fully the following tasks. We already started to consistently generate kinetic coefficients using multiscale approach using the Lammps package. Assessment of the current state of drinking water supply to settlements of the Republic of Kazakhstan; the current state of water treatment facilities for the purification of natural waters; build a new design of water intake and treatment facilities; theory, experiments of the technology of slow filtration at low-capacity facilities for the purification of natural waters; theory and experiments for the development of modified sorbents from natural minerals of Kazakhstan, for the purification of natural waters. Development of natural water purification technology using a complex of sorbent and coagulants makes Comsol models attractive which includes the treatment of sludge from waterworks, ensuring their utilization.